\documentclass[lettersize,journal]{IEEEtran}
\usepackage{amsmath,amsfonts}
\usepackage{algorithmic}
\usepackage{algorithm}
\usepackage{array}
\usepackage{multirow}
\usepackage[caption=false,font=normalsize,labelfont=sf,textfont=sf]{subfig}
\usepackage{textcomp}
\usepackage{stfloats}
\usepackage{url}
\usepackage{verbatim}
\usepackage{graphicx}
\usepackage{subfig}
\graphicspath{{figure/}}
\usepackage{cite}
\def\BibTeX{{\rm B\kern-.05em{\sc i\kern-.025em b}\kern-.08em
		T\kern-.1667em\lower.7ex\hbox{E}\kern-.125emX}}

\begin{document}

\title{LEOSTP: A Spatio-Temporal Traffic Prediction Framework for LEO Satellite Networks}

\author{Shaoyou Ao,~\IEEEmembership{Graduate~Student~Member,~IEEE},
	Yong~Niu,~\IEEEmembership{Senior~Member,~IEEE},
	Zhu~Han,~\IEEEmembership{Fellow,~IEEE},
	Cheng~Li,~\IEEEmembership{Member,~IEEE},
	and Bo~Ai,~\IEEEmembership{Fellow,~IEEE}

\thanks{S. Ao, Y. Niu (\textit{Corresponding author}), C. Li, B. Ai are with the School of electronic and Information Engineering and  the Beijing Engineering Research Center of High-speed Railway Broadband Mobile Communications, Beijing Jiaotong University, Beijing 100044, China.}

\thanks{Z. Han is with the Department of Electrical and Computer Engineering at the University of Houston, Houston, TX 77004 USA, and also with the Department of Computer Science and Engineering, Kyung Hee University, Seoul, South Korea, 446-701.}
}



\maketitle

\begin{abstract}
	With the evolution of next-generation mobile communication networks and the commercial boom of Low Earth Orbit (LEO) satellites, globally covered satellite networks are gradually becoming a crucial infrastructure for massive user access and seamless connectivity. 
	Accurate traffic prediction is crucial for maintaining the quality of service (QoS) and resource allocation efficiency in satellite networks. 
	However, existing methods struggle to effectively address the three major challenges of LEO networks: highly complex temporal dynamics caused by satellite cross-regional movement, multivariate dependencies in multi-satellite collaboration, and strong spatial heterogeneity driven by user distribution, human activity intensity, and local geographic environments. 
	In this article, we propose a LEO Satellite Traffic Predictor (LEOSTP) framework, a diffusion model-based end-to-end model that forecasts future satellite traffic by jointly leveraging historical traffic patterns and contextual characteristics of the corresponding service regions. 
	The framework consists of two core modules: 1) The general traffic feature extractor module combines the diffusion process with a Transformer architecture to model the multi-scale temporal features of the traffic itself.  
	2) The external condition encoder module integrates geographic semantic information such as population distribution, point-of-interest (POI) distribution, and local time into the prediction process through a Transformer-based encoder. 
	In this way, the model captures the deep correlation between the external environment and traffic dynamics. 
	Experimental results based on large-scale simulated constellation data show that LEOSTP significantly outperforms traditional statistical models such as ARIMA and SVR, and classical sequence models including LSTM and Transformer, in prediction accuracy.
	
\end{abstract}


\section{Introduction}
\IEEEPARstart{O}ver the past several decades, with the continuous evolution of global mobile communication networks, people have placed higher expectations on ubiquitous broadband access. 
Enabled by advances such as low-cost satellite manufacturing, reusable launch vehicles, and rapid progress in space-based communication technologies, Low Earth Orbit (LEO) satellite networks are emerging as a pivotal infrastructure for future ubiquitous wireless access. 
These developments position LEO constellations not only as a complementary layer to terrestrial networks, but also as a critical component in delivering seamless, wide-area, and resilient broadband services worldwide. 
Accurately forecasting traffic demand on satellite links is becoming increasingly vital for delivering stable and high-quality satellite Internet services. 
For example, if the network scheduling system can anticipate variations in the service load of each satellite over a future period, it can proactively plan inter-satellite forwarding and dynamically allocate network resources. 
In addition, advance traffic awareness enables more effective scheduling of ground gateway stations, thereby avoiding congestion in hotspot areas and improving overall quality of service (QoS) and resource utilization efficiency \cite{wu2025sate, 11015330}. 
More importantly, accurate traffic and load prediction has direct implications for practical satellite control and scheduling. 
Since LEO satellites spend much of their orbital period serving low-demand remote regions and only briefly traverse high-activity areas, advance knowledge of traffic demand enables region-aware control strategies, such as conservative resource use in sparse areas and proactive preparation for peak loads \cite{9520323}. 
Similar to how traffic prediction supports energy-efficient operations in terrestrial networks, accurate satellite traffic forecasting serves as an upstream decision input for routing, resource allocation, and network scheduling, making it essential for efficient LEO network operation. 

Existing research typically treats network traffic prediction as a general time-series forecasting problem and analyzes it using statistical or deep learning methods. 
To improve the accuracy of traffic forecasting, extensive work has been conducted along this line. 
Traditional methods such as Seasonal Autoregressive grated Moving Average (SARIMA) and Support Vector Regression (SVR) exhibit a certain degree of nonlinear fitting capability, but they struggle to capture spatial correlations, which are both common and crucial in mobile networks \cite{JIANG2022117163}. 
Many researchers have also employed machine learning techniques to model traffic prediction problems. To overcome the limitations of traditional methods, recurrent neural network (RNN) and long short-term memory network (LSTM) have been introduced to capture the distribution and trends of cellular traffic sequences \cite{8553663,10071958, 10693779}. 
Meanwhile, researchers have recognized the spatial correlation in satellite traffic and adopted methods such as graph convolutional networks (GCNs) to incorporate graph-structured data and exploit the spatial dependencies between nodes, thereby enhancing the predictive performance of the model \cite{zhang2025self,9801679, 10978666, 11152704}. 
In \cite{10938839}, a spatio-temporal graph attention and gated convolutional network prediction model based on spatio-temporal feature extraction (ST-GAGCN) is proposed, which extracts the most critical features in LEO satellite traffic prediction, thereby improving its performance. 
These models integrate temporal and spatial forecasting approaches, using time-series analysis to capture dynamic patterns of traffic flow over time, while also applying spatial analysis methods to reveal traffic distribution characteristics across different geographical locations. 
However, in terms of leveraging spatial information, most methods adopt a gridded approach to partition geographical regions and rely on the model to extract spatial correlations, which does not reflect the inherent differences between spatial locations.

\begin{figure*}[!ht]
	\begin{center}
		\includegraphics*[width=1.9\columnwidth]{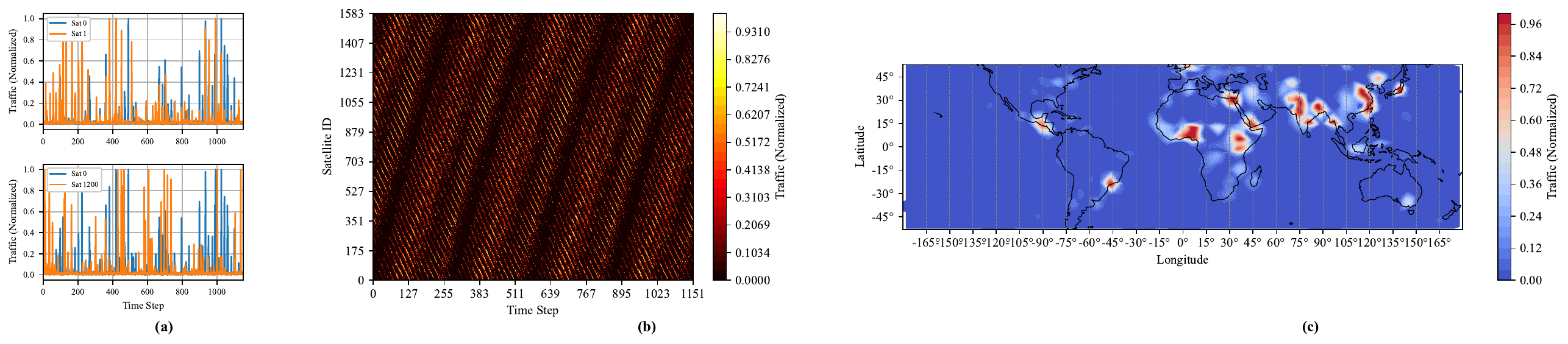}
	\end{center}
	\caption{Multi-perspective visualization of LEO satellite traffic characteristics: a) Typical traffic time series of multiple satellites; b) Global traffic heatmap across all satellites; c) Geographical distribution of normalized mobile traffic at a single time snapshot}
	\label{fig_feature}
\end{figure*}

In this paper, we propose the LEO Satellite Traffic Predictor (LEOSTP), an end-to-end traffic prediction framework based on a diffusion model, to forecast future traffic for LEO satellites. 
LEOSTP first models geographic semantic features from a spatial perspective using an external conditional encoder, uniformly encoding multiple external information such as the Points of Interest (POI) distribution, local time, and population grids into a compact conditional tensor to represent potential demand patterns within the satellite coverage area. 
Subsequently, the traffic prediction module, built upon a conditional diffusion model, integrates external conditions and historical traffic during the reverse denoising process, enabling the model to simultaneously capture both area switching caused by orbital migration and traffic fluctuations driven by user activity. 
By leveraging this joint modeling approach, LEOSTP achieves higher accuracy in satellite traffic prediction under environments characterized by multi-scale temporal dependencies, dynamic spatial structures, and complex geographic semantics.

The contributions of this article can be summarized as:
\begin{itemize}[]
	\item We propose LEOSTP, an end-to-end diffusion-based framework for LEO satellite traffic prediction, which effectively captures spatio-temporal patterns and dynamic geographic semantics by integrating external condition information such as population distribution, POI distribution, and local time.
	\item We demonstrate significant performance gains enabled by external condition modeling, showing that LEOSTP improves regional demand perception and achieves a 15.91\% performance improvement over existing methods on large-scale simulated satellite datasets constructed with probabilistic user behavior modeling and region-dependent activity rhythms. 
\end{itemize}

The remainder of this article is organized as follows. 
In Section \ref{sec2}, we introduce the core challenges of satellite traffic forecasting and provide a visual analysis of the spatio-temporal characteristics of satellite traffic data. 
In Section \ref{sec3}, we present the proposed LEOSTP framework and describe the design of its key modules. 
Then, in Section \ref{sec4}, we evaluate the model using a large-scale simulation dataset and compare its performance with several state-of-the-art methods.
Finally, in Section \ref{sec5}, we conclude the paper and discuss potential directions for future research.

\section{Challenges and Overview}\label{sec2}

In communication networks, mobile traffic refers to the volume of data transmitted between a base station and a mobile terminal over a wireless channel within a given time interval. 
In satellite networks, this concept naturally extends to the traffic carried by satellites while serving ground users within their coverage areas, which varies over time as satellites move along their orbits. 
For satellite mobile traffic, as a spatio-temporal series, its prediction faces key challenges similar to those of traditional time-series forecasting in terms of sequence information modeling. 
In addition, the traffic behavior of LEO satellite networks exhibits unique characteristics that differ entirely from those of terrestrial cellular networks, thus making it difficult to directly transfer traditional methods. 
Specifically, there are three major challenges:

\textbf{Challenge 1: Complex Temporal Variation. }As shown in Fig. \ref{fig_feature}(a), satellite traffic time series exhibit significant bursts and high-frequency fluctuations, and their statistical characteristics change continuously over time, reflecting strong non-stationarity. 
Unlike terrestrial cellular traffic, whose periodicity is largely governed by stable diurnal cycles within fixed service areas, LEO satellite traffic exhibits multi-scale periodic components. 
These components are jointly driven by the Earth's rotation, orbital repetition periods, and user activity rhythms, and are repeatedly interrupted by coverage migration. 
As a result, the traffic pattern becomes a hybrid, where peak loads emerge over densely populated regions or transportation routes, while traffic drops sharply over oceans or sparsely populated areas. 
This complex temporal variability makes traditional forecasting methods based on stationary or fixed-period assumptions insufficient. 

\textbf{Challenge 2: Rapidly Reconstructed Spatial Correlations Across Moving Satellites. }Fig. \ref{fig_feature}(b) shows that the spatial structure of the LEO constellation changes rapidly as satellites move, causing adjacency relationships to be continuously reconstructed. 
Satellites in the same orbit successively scan identical geographical regions, forming a cross-time and cross-satellite coupling structure that appears as a diagonal pattern in the heatmap. 
In contrast, satellites in different orbital planes exhibit substantially weaker correlations due to divergent service paths, highlighting the heterogeneity of spatial dependencies. 
Because of this constantly shifting topology, models that rely on fixed spatial graphs or static neighborhood structures cannot effectively capture the evolving spatial correlations. 
In essence, the key difficulty lies not in the existence of spatial correlations, but in the fact that such correlations are continuously reconstructed as satellites move along their orbits.

\textbf{Challenge 3: Semantic Mismatch Between Fixed Models and Migrating Coverage Areas. }Beyond the dynamic reconstruction of spatial correlations, a more fundamental challenge arises from the continuous migration of coverage semantics. 
Fig. \ref{fig_feature}(c) illustrates the coverage area migration phenomenon, where satellites serve entirely different regions over time. 
The service demands in these regions vary with factors such as population distribution, local time, and human activity intensity, resulting in pronounced spatial heterogeneity and temporal asynchrony. 
As a result, similar traffic magnitudes may arise from fundamentally different underlying causes, such as satellite coverage over dense urban regions or transient high-activity periods. This semantic ambiguity is often implicitly overlooked in traditional terrestrial traffic prediction models. 
Unlike terrestrial cellular networks with stable service regions, LEO networks require models that can continuously adapt to semantic shifts in coverage areas. 
Traditional approaches that depend on static geographic labels or region-specific features therefore experience severe mismatch when the underlying semantics change. 

These observations collectively indicate that satellite traffic patterns are not solely governed by temporal continuity or inter-satellite adjacency, but are fundamentally driven by the semantic characteristics of the underlying coverage areas. 
This insight suggests the need to move beyond purely sequence-based or topology-based modeling and explicitly incorporate geographic semantic information as a guiding prior for traffic prediction.

\begin{figure}[!t]
	\begin{center}
		\includegraphics*[width=1.0\columnwidth]{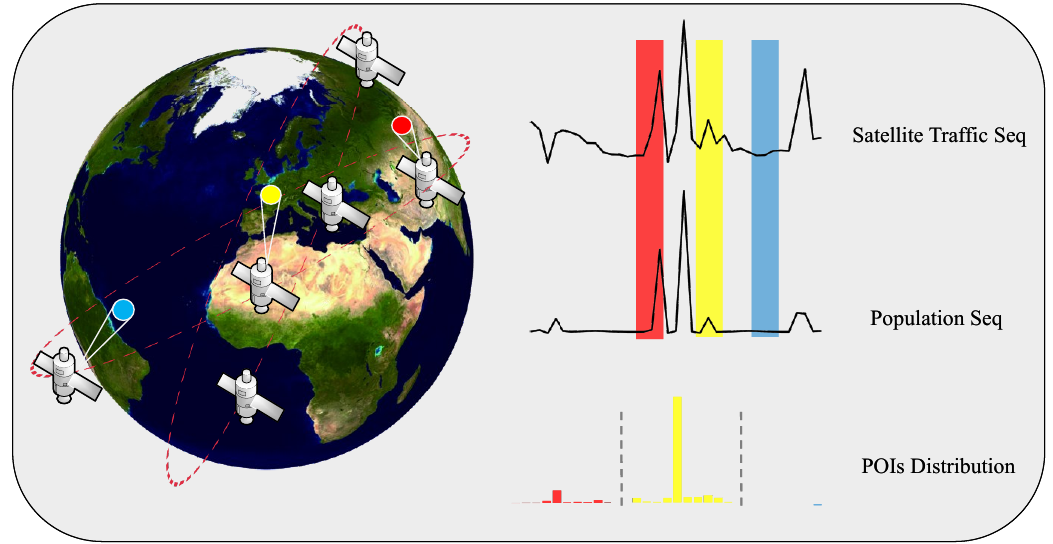}
	\end{center}
	\caption{Illustration of the spatio-temporal dynamics in LEO satellite traffic prediction.}
	\label{fig_motivation}
\end{figure}

Fig. \ref{fig_motivation} illustrates the core motivation behind our work and highlights why existing forecasting approaches struggle in LEO satellite networks. 
Traditional time series forecasting methods primarily rely on historical traffic sequences of individual satellites and are effective in terrestrial cellular networks with fixed coverage areas and stable service targets. 
However, for LEO satellites whose coverage regions migrate continuously at high speeds, historical sequences alone are insufficient to characterize rapidly changing service demands.

Some studies attempt to introduce spatial information by jointly modeling traffic from a fixed set of neighboring satellites to capture spatial correlations \cite{10970731}. 
Nevertheless, such approaches remain limited, as satellites sequentially traverse regions with vastly different population densities, local time profiles, human activity patterns, and POI distributions. 
These geographic semantics lead to substantial variations in traffic demand that cannot be adequately captured through neighboring satellite traffic alone. 
Motivated by this observation, we explicitly model external geographic semantics by encoding multi-source auxiliary information such as population, local time, and POI distributions as conditional features, enabling more accurate traffic prediction under coverage migration. 
The detailed model structure is presented in subsequent sections.

\section{Model and Solution}\label{sec3}
\begin{figure}[!t]
	\begin{center}
		\includegraphics*[width=0.9\columnwidth]{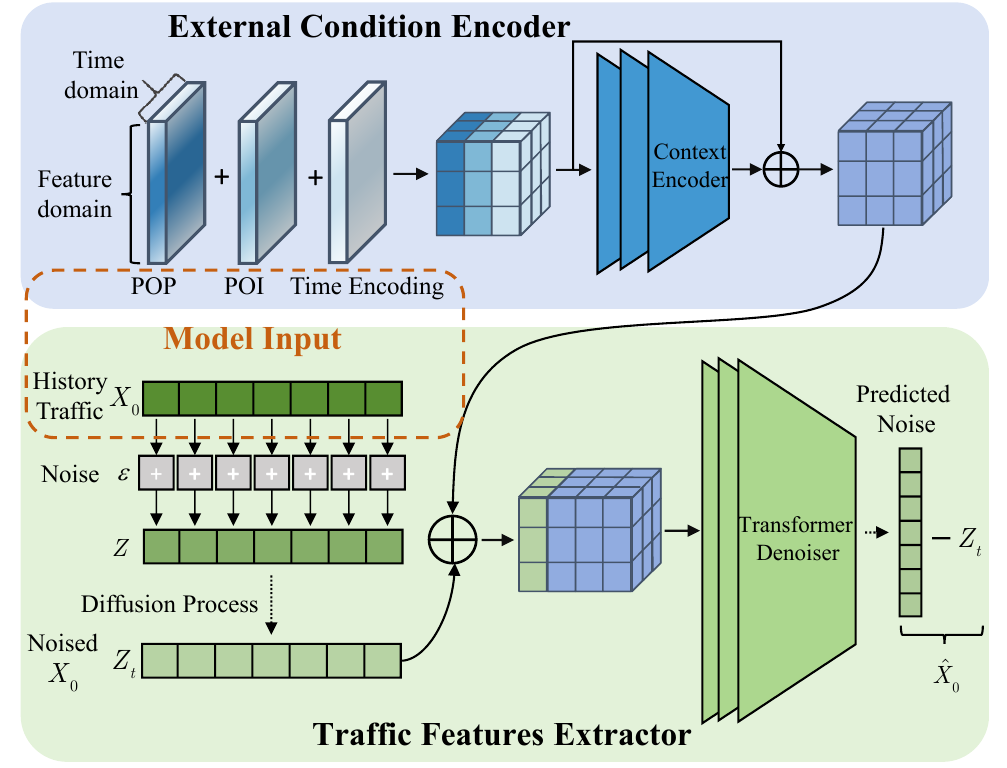}
	\end{center}
	\caption{Overall architecture of the proposed LEOSTP framework.}
	\label{fig_framework}
\end{figure}

Fig. \ref{fig_framework} illustrates the overall architecture of our proposed LEOSTP, an end-to-end LEO satellite traffic prediction framework based on a conditional diffusion model. 
Its core idea is to explicitly decouple and fuse the temporal characteristics of the satellites with the geographic semantics of the coverage areas, allowing the model to capture traffic variations driven by both orbital dynamics and geographic demand. 
The overall framework consists of two parallel feature channels: the External Condition Encoder and the Traffic Features Extractor. 

The model input consists of two components: historical traffic sequences, which include the main traffic sequence of the target satellite and auxiliary sequences from orbitally related neighboring satellites, and discrete geographic conditions, including population density, POI distributions, surface attributes, and local time, which jointly determine the potential service demand within the satellite coverage area. 
In the external condition encoder, all geographic factors are first mapped into a unified feature domain, after which their temporal semantic dependencies are extracted through a context encoder to generate compact geographic semantic vectors that characterize the demand intensity and functional attributes of the coverage area. 

Meanwhile, historical traffic sequences are fed into a traffic feature extractor. 
This module adds noise to the input sequence according to the diffusion process, after which a Transformer-based denoising network gradually recovers its underlying structure to extract key temporal trends and periodic patterns. 
To ensure that the denoising process is aware of spatial semantics, the encoded geographic condition vector is fused with the temporal features and injected as conditional input into each step of the reverse diffusion process. 

Unlike conventional regression-based forecasting methods that directly map historical inputs to point estimates, diffusion models formulate traffic prediction as a progressive generative process. 
This formulation is particularly suitable for LEO satellite traffic, where sudden demand surges, coverage transitions, and stochastic user behaviors introduce substantial uncertainty. 
By iteratively denoising traffic representations under the guidance of external conditions, the diffusion process enables the model to flexibly adapt its prediction trajectory to changing geographic semantics, rather than enforcing a fixed deterministic mapping.

Ultimately, the conditional diffusion model, guided jointly by external geographical conditions and temporal information, recursively predicts future loads step by step. 
Through this structured joint modeling approach, LEOSTP can effectively capture spatial coverage changes caused by satellite orbital migration and the strong non-stationarity introduced by varying human activity patterns across different regions, thereby achieving more accurate predictions of future loads. 
The details of our model are described as follows. 

\subsection{Traffic Features Extractor}
To address the challenge of complex temporal variation caused by orbital dynamics and coverage migration, we design a Traffic Features Extractor to capture multi-scale temporal dependencies from noisy traffic sequences. 
This module is built on a Transformer-based diffusion structure and incorporates conditional diffusion mechanisms, enabling the model to exploit both historical traffic features and external conditional information during denoising.

First, we uniformly embed the input multimodal traffic sequences to obtain stable token representations, which facilitates the capture of sequence structure during the diffusion inversion process. 
We then employ a Transformer as the denoising backbone, allowing the model to simultaneously perceive long-term trends, orbital periodicity, and short-term fluctuations within an extended sequence window. 
The Transformer enhances the features of noise-contaminated sequences and enables the network to gradually approximate the true traffic distribution during reverse diffusion, thereby achieving high-precision reconstruction of future traffic.

\begin{table*}[!t]
	\centering
	\caption{Summary of the Satellite Traffic and External Condition Dataset}
	\label{table1}
	
	\renewcommand{\arraystretch}{1.5}
	\setlength{\tabcolsep}{4pt}
	
	\begin{tabular}{|c|c|c|c|c|c|}
		\hline
		\textbf{Category} & \multicolumn{5}{c|}{\textbf{Description}} \\
		\hline
		
		\textbf{LEO Constellation} 
		& 1584 satellites
		& 24 planes 
		& 550 km altitude 
		& 53$^\circ$ inclination 
		& 15$^\circ$ visibility \\
		\hline
		
		\textbf{Traffic Data} 
		& \multicolumn{2}{c|}{96 h (Started at 00:00 UTC on 1/1/2025)}
		& 5-minute interval
		& 1152 time steps
		& $>$75,000 sequences \\
		\hline
		
		\textbf{Population Data} 
		& \multicolumn{5}{c|}{LandScan Global 2023 (1 km resolution)} \\
		\hline
		
		\textbf{POI Categories} 
		& \multicolumn{5}{c|}{%
			\parbox[c]{12cm}{\centering 
				\rule{0pt}{1.2em}
				10 categories from OSM (Residential, Commercial, Industrial, Education, Transport, Government,
				Tourism, Healthcare, Green, Water)
		}} \\
		\hline
	\end{tabular}
\end{table*}

\subsection{External Condition Encoder}
To mitigate the semantic mismatch introduced by continuously migrating coverage areas, we construct an External Condition Encoder that explicitly models dynamic geographic and social semantics as conditional priors for traffic prediction. 
This module uniformly encodes multi-source external conditions and injects them as conditioning information into the diffusion process. 
Built on a Transformer architecture, it captures the temporal variation patterns of external features and extracts cross-regional and cross-temporal contextual dependencies. 
The inputs include population distribution, POI distribution, and local time information corresponding to the satellite coverage area, which collectively determine potential service demand levels. 
After processing by the Transformer, these external factors are mapped into condition vectors aligned with temporal traffic features. 
Unlike terrestrial networks with fixed coverage, the coverage areas of LEO satellites continuously migrate with their orbits; therefore, external features must be modeled as dynamic sequences rather than static attributes. 
This encoder ensures that the conditional features accurately reflect temporal changes in spatial and social semantics and guides the inversion direction of the Traffic Features Extractor during the diffusion denoising process, allowing predictions to better match the geographic semantic–driven traffic dynamics.

In summary, LEOSTP consists of two core modules specifically designed for satellite traffic prediction. 
The Traffic Features Extractor is responsible for recovering temporal structure from noisy sequences, and the External Condition Encoder provides dynamic conditional priors across spatial, social, and temporal dimensions. 
Together, these modules form the foundation of conditional diffusion–based prediction, enabling the model to maintain high robustness and interpretability even in scenarios with strong non-stationarity and continuous coverage migration. 
In the next section, we evaluate the predictive performance of the proposed framework using simulated constellation data. 

\section{Evaluation on Satellite Traffic Data Dataset}\label{sec4}
In the experiment, we used a simulated LEO satellite constellation traffic dataset to comprehensively evaluate the performance of the proposed model. 
The dataset contains 1584 satellites at an orbital altitude of 550 km and an inclination of 53$^\circ$, evenly distributed across 24 orbital planes. 
The simulation was conducted from 00:00 UTC on January 1, 2025, to 00:00 UTC on January 5, 2025, covering a total of 96 hours with a sampling interval of 5 minutes. 

We generate a large-scale simulated LEO satellite traffic dataset by following the traffic modeling framework proposed in \cite{wu2025sate}. Specifically, the Earth’s surface is discretized into latitude–longitude grids with a spatial resolution of 1 degree. 
Based on global population distribution data from the LandScan Global Dataset (2023) with a spatial resolution of 1 km, approximately 10 million users are probabilistically placed across these grids to reflect realistic spatial heterogeneity in human presence and activity intensity. 

To model temporal variations in user activity, POI semantics are incorporated using data extracted from OpenStreetMap (OSM), an open-source geographic database. 
Ten categories of POIs are considered, including residential, commercial, and industrial areas, transportation hubs, as well as functional facilities such as medical, educational, and tourism locations.
For each POI category, multiple Gaussian distributions are superimposed to generate distinct diurnal activity curves, capturing the time offsets of peak user activity across different functional regions. 
In addition, local time information is used to construct time-dependent user activation probability curves, enabling probabilistic modeling at both the user activation and service arrival stages. 
This multi-level stochastic modeling improves the realism of the simulated human activity dynamics.

Based on these activation probabilities, users are dynamically activated according to the modeled activity patterns. 
Each activated user generates traffic within each sampling interval following a probabilistic service model that includes three representative traffic types: voice services at 64 kbps, video services at 10 Mbps, and file transfer services at 50 Mbps. 
As satellites traverse the Earth along their orbital trajectories, they sample and aggregate traffic from users located within their instantaneous coverage areas. 
Through this process, more than 75,000 satellite traffic sequences are generated, forming the final dataset used for model training and evaluation. 
The overall structure of the dataset is summarized in Table \ref{table1}.

To evaluate the predictive performance of the proposed model, we compared LEOSTP with several representative time series forecasting methods, including traditional statistical models, classic machine learning methods, and deep learning baseline models:
\begin{itemize}
	\item \textbf{ARIMA:} As a widely used traditional time series forecasting method, ARIMA transforms non-stationary sequences into stationary forms through differencing and fits them using autoregressive and moving average components. In our experiment, ARIMA is implemented using the Python statsmodels library. 
	
	\item \textbf{SVR}: A classic machine learning regression approach, SVR performs prediction by mapping features into a high-dimensional space and identifying a regression hyperplane with the maximum margin under an allowable error threshold. We implement SVR using the Python scikit-learn. 
	
	\item \textbf{LSTM:} Capable of capturing long-term dependencies in time series, LSTM is suitable for complex nonlinear temporal data and is widely applied in various sequence forecasting tasks. 
	 
	\item \textbf{Transformer:} By employing a self-attention mechanism, the Transformer models global dependencies between arbitrary positions in a sequence, making it particularly effective for long-sequence modeling and multivariate coupled forecasting scenarios. 

\end{itemize}

Building upon these baselines, our LEOSTP model adopts an end-to-end conditional diffusion structure, enabling generative prediction of future sequences without manual feature engineering. 
The diffusion denoising network is optimized using mean squared error (MSE) as the objective function and trained with the Adam optimizer with an initial learning rate of $1\times10^{-3}$. 
To enhance generalization, we apply regularization techniques including a Dropout rate of 0.1, L2 weight decay of $1\times10^{-5}$, and a cosine annealing learning rate schedule with a minimum value of $1\times10^{-6}$. 
The model is implemented using a Transformer architecture with 64 hidden dimensions, 8 attention heads, and a 3-layer encoder. 
The training batch size is set to 256, and the model is trained for 300 epochs with an early stopping strategy that uses a patience value of 20 to mitigate overfitting. 
The dataset is divided into training, validation, and test sets in an 8:1:1 ratio, and the normalized root mean square error (NRMSE) is used as the primary evaluation metric. 

\begin{figure}[!t]
	\begin{center}
		\includegraphics*[width=0.9\columnwidth]{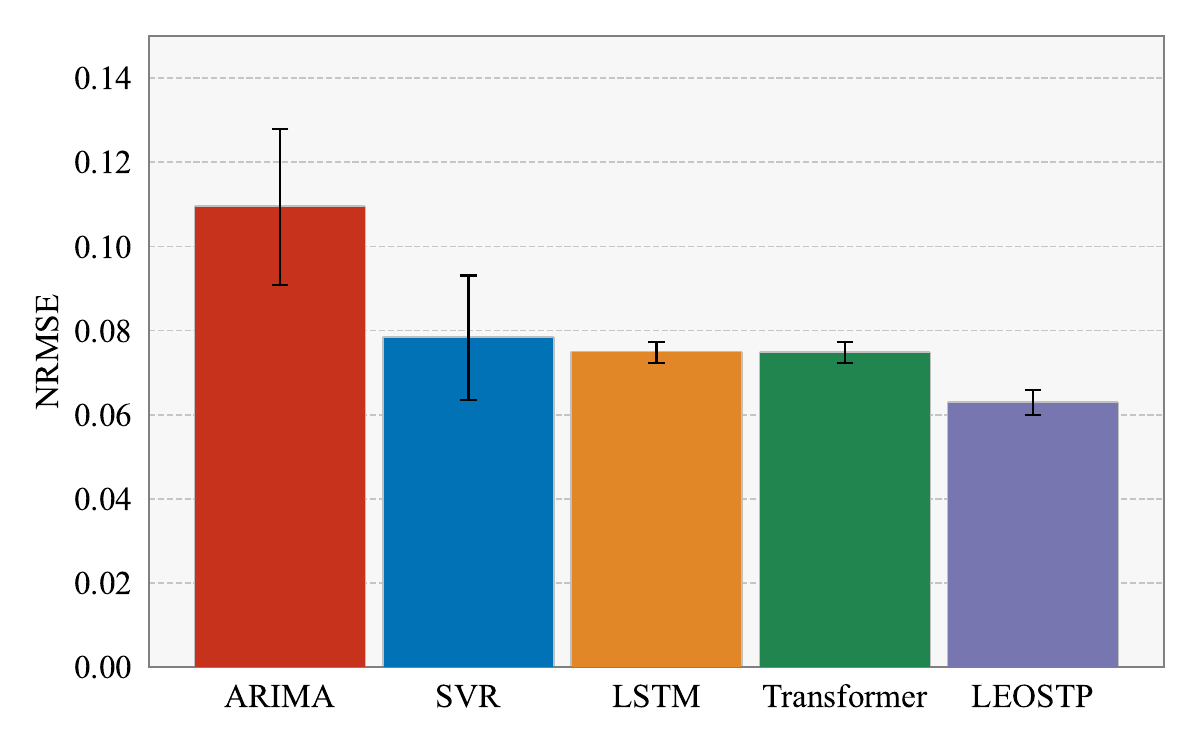}
	\end{center}
	\caption{Performance comparison between LEOSTP and baselines.}
	\label{fig_simulation}
\end{figure}

\subsection{Prediction Results}
In this section, we deploy LEOSTP on the test set and compare it with four representative benchmark methods. 
As shown in Fig. \ref{fig_simulation}, ARIMA yields the largest prediction error, with an NRMSE above 0.10, indicating that classical linear statistical modeling is insufficient for capturing the highly non-stationary and nonlinear nature of LEO satellite traffic. SVR performs substantially better than ARIMA, with an NRMSE around 0.08, benefiting from kernel-based nonlinear regression, but it still exhibits relatively large variance as reflected by the error bar. 

Deep sequence models such as LSTM and Transformer provide moderate improvements over traditional methods, but their error reduction remains limited without explicit modeling of geographic semantics and external factors. After incorporating external condition information, LEOSTP achieves the best performance on the test set, reaching the lowest NRMSE (around 0.06). Compared with the best-performing sequence-based baseline, LEOSTP achieves a 15.91\% reduction in prediction error. The smaller error bar also indicates improved stability and robustness across samples. 
This indicates that the external condition encoder successfully captures exogenous factors such as population size, functional attributes, and diurnal rhythms within the coverage area, providing strong conditional guidance for diffusion-based prediction and enabling the model to more accurately characterize the traffic evolution process. 

\begin{figure}[!t]
	\begin{center}
		\includegraphics*[width=0.9\columnwidth]{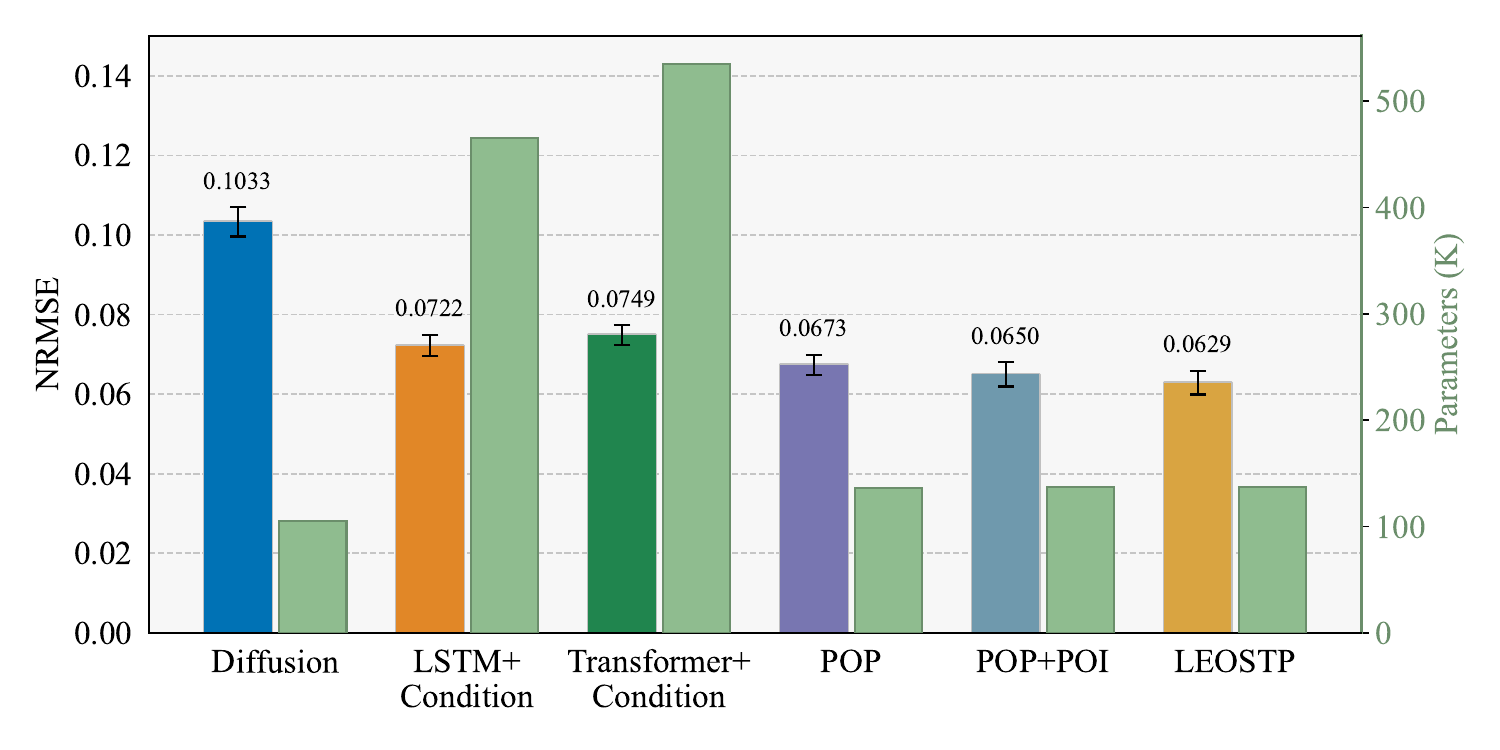}
	\end{center}
	\caption{Ablation results of different external conditions.}
	\label{fig_ablation}
\end{figure}

\subsection{Ablation Study}
To further analyze the impact of external conditions on model performance, we conduct an extended ablation study that includes both feature-level and model-level comparisons. 
Specifically, we evaluate three types of conditional features used in LEOSTP, namely population distribution (POP), POI distribution, and time encoding, and further introduce two conditional sequence baselines, where external conditions are directly incorporated into LSTM and Transformer models for auxiliary prediction. 

Fig. \ref{fig_ablation} presents the prediction results under different feature and model configurations, where the horizontal axis indicates the configuration and the vertical axes report NRMSE and the number of model parameters, respectively. 
The results show that removing all conditional features leads to a significant performance degradation, confirming that relying solely on historical traffic sequences is insufficient for high-precision satellite traffic forecasting. 
Among individual features, POP provides the most significant performance gain, as it directly reflects the scale and intensity of human activity and serves as the primary driver of traffic demand. The contribution of POI features is relatively limited because most LEO satellite trajectories cover low-activity regions with sparse POIs, although POIs provide useful complementary information when satellites traverse high-activity urban areas. 
Time encoding yields only modest improvement, since periodic temporal patterns such as diurnal rhythms can be implicitly learned by sequence models, making explicit time features mainly supportive rather than dominant.

We further observe that simply injecting external conditions into conventional LSTM and Transformer models yields only marginal performance gains, despite increasing model complexity. In comparison, LEOSTP achieves superior accuracy with a more parameter-efficient integration of external information. Overall, the results validate the importance of explicitly modeling geographic semantics as conditional priors within the diffusion framework, rather than relying on naïve feature concatenation.

\subsection{Computational Complexity and Deployment Considerations}
The traffic prediction task addressed in this work operates at the network layer, where traffic sequences are typically modeled at a minute-level granularity and thus tolerate higher computational latency than physical-layer processing. 
Moreover, such models are deployed on satellite platforms or network controllers with substantially greater computational resources than mobile terminals. Our ablation study shows that LEOSTP does not introduce a significant increase in model parameters compared with existing deep learning baselines. 
Although diffusion-based models involve iterative denoising and incur higher computational overhead, this cost is acceptable for latency-insensitive traffic forecasting. 
In practice, LEOSTP achieves an inference time within 1.5 ms on an RTX 4070 Ti GPU, satisfying practical deployment requirements. 
Furthermore, traffic prediction has already been adopted in terrestrial networks to support tasks such as base station deployment planning and dynamic sleeping, suggesting that accurate satellite traffic forecasting can similarly serve as an effective upstream decision-support module for load-aware control and scheduling in LEO networks. 

\section{Conclusion}\label{sec5}
This paper proposes LEOSTP, an end-to-end conditional diffusion traffic prediction framework for LEO satellite networks. 
Unlike traditional methods that rely solely on the historical sequences of a single satellite, LEOSTP simultaneously models three key factors: multi-scale temporal features, orbital dynamic spatial dependencies, and regional geographic semantics, thereby achieving a comprehensive characterization of LEO satellite traffic prediction. 
Specifically, the Traffic Features Extractor captures the intrinsic temporal structure of satellite traffic, while the External Condition Encoder incorporates essential external conditions such as population distribution, POI layout, and local time. 
These two components jointly guide the prediction process during diffusion inversion, enabling the model to effectively adapt to the strong non-stationarity caused by rapidly shifting satellite coverage areas. 
Experimental results based on large-scale simulated constellation data show that LEOSTP significantly outperforms representative baseline models such as ARIMA, SVR, LSTM, and Transformer in terms of NRMSE, achieving an overall performance improvement of 15.91\% and demonstrating clear performance advantages. 
Ablation studies further highlight the importance of external conditions: population and temporal features make the greatest contribution to prediction accuracy, while POI information, although limited by geographic sparsity, enhances the model’s sensitivity to sudden traffic surges when satellites traverse high-activity regions. 
Overall, LEOSTP provides a scalable and efficient solution for refined traffic forecasting in complex and dynamic LEO constellation environments. In future work, we will explore integrating real-world measurement data, higher-dimensional multimodal geographic semantics, and online adaptive forecasting mechanisms to further advance intelligent and autonomous traffic management in LEO satellite networks.

\bibliographystyle{IEEEtran}
\bibliography{references}

@inproceedings{wu2025sate,
	title={{SaTE}: Low-Latency Traffic Engineering for Satellite Networks},
	author={Wu, Hao and Han, Yizhan and Rajpal, Mohit and Zhang, Qizhen and Wang, Jingxian},
	booktitle={Proceedings of the ACM SIGCOMM Conference},
	address = {S\~{a}o Francisco Convent, Coimbra, Portugal},
	pages={896-916},
	year={2025},
	month={Sep.},
	volume={},
	number={},
}

@article{JIANG2022117163,
	title = {Cellular traffic prediction with machine learning: A survey},
	journal = {Expert Systems with Applications},
	volume = {201},
	pages = {117163},
	year = {2022},
	month ={Apr.},
	issn = {0957-4174},
	doi = {https://doi.org/10.1016/j.eswa.2022.117163},
	author = {Weiwei Jiang},
}

@ARTICLE{9520323,
	author={Wu, Hao and Yan, Jian},
	journal={IEEE Network}, 
	title={{QoS} Provisioning in Space Information Networks: Applications, Challenges, Architectures, and Solutions}, 
	year={2021},
	month={Jul.},
	volume={35},
	number={4},
	pages={58-65},
	doi={10.1109/MNET.011.2100056}}

@ARTICLE{10938839,
	author={Chen, Chen and Sun, Chengxin and Li, Huimin and Jin, Fan and Pei, Qingqi and Wan, Shaohua},
	journal={IEEE Transactions on Aerospace and Electronic Systems}, 
	title={{ST-GAGCN-LEO}: A Spatiotemporal Graph Attention and Gated Convolutional Network for {LEO} Satellite Traffic Prediction}, 
	year={2025},
	month={Aug.},
	volume={61},
	number={4},
	pages={9669-9685},
	doi={10.1109/TAES.2025.3552548}}

@ARTICLE{10071958,
	author={Kavehmadavani, Fatemeh and Nguyen, Van-Dinh and Vu, Thang X. and Chatzinotas, Symeon},
	journal={IEEE Transactions on Wireless Communications}, 
	title={Intelligent Traffic Steering in Beyond {5G} Open {RAN} Based on {LSTM} Traffic Prediction}, 
	year={2023},
	month={Nov.},
	volume={22},
	number={11},
	pages={7727-7742},
	doi={10.1109/TWC.2023.3254903}}

@article{zhang2025self,
	title={Self-similar traffic prediction for {LEO} satellite networks based on {LSTM}},
	author={Zhang, Yan and Wang, Yong and Cao, Haotong and Hu, Yihua and Lin, Zhi and An, Kang and Li, Dong},
	journal={IET communications},
	volume={19},
	number={1},
	pages={e12863},
	year={2025},
	month={Jan.},
	publisher={Wiley Online Library}
}

@ARTICLE{8553663,
	author={Feng, Jie and Chen, Xinlei and Gao, Rundong and Zeng, Ming and Li, Yong},
	journal={IEEE Network}, 
	title={{DeepTP}: An End-to-End Neural Network for Mobile Cellular Traffic Prediction}, 
	year={2018},
	month={Nov.},
	volume={32},
	number={6},
	pages={108-115},
	doi={10.1109/MNET.2018.1800127}}

@INPROCEEDINGS{10693779,
	author={Gong, Lizeng and Chen, Quan and Yang, Lei and Yin, Zhenglong and Yang, Huaguo and Li, Jiaqi},
	booktitle={IEEE/CIC International Conference on Communications in China (ICCC Workshops)}, 
	title={Satellite Coverage Traffic Deviation Prediction for {LEO} Constellation Networks Based on {CNN-LSTM} Method}, 
	pages={681-686},
	year={2024},
	month={Aug.},
	address={Hangzhou, China},
	volume={},
	number={},
	doi={10.1109/ICCCWorkshops62562.2024.10693779}}

@ARTICLE{9801679,
	author={Wang, Zi and Hu, Jia and Min, Geyong and Zhao, Zhiwei and Chang, Zheng and Wang, Zhe},
	journal={IEEE Transactions on Industrial Informatics}, 
	title={Spatial-Temporal Cellular Traffic Prediction for {5G} and Beyond: A Graph Neural Networks-Based Approach}, 
	year={2023},
	month={Apr.},
	volume={19},
	number={4},
	pages={5722-5731},
	doi={10.1109/TII.2022.3182768}}

@INPROCEEDINGS{10978666,
	author={Zhu, Zihan and Wang, Yuzhuo and Wu, Ke and Hou, Yunpeng and Hel, Huasen and Yang, Jian},
	booktitle={IEEE Wireless Communications and Networking Conference (WCNC)}, 
	title={A Novel Traffic Prediction Method for Dynamic Satellite Networks Based on Graph Attention Networks}, 
	year={2025},
	month={Mar.},
	volume={},
	number={},
	pages={},
	address={Milan, Italy},
	doi={10.1109/WCNC61545.2025.10978666}}

@ARTICLE{11015330,
	author={Liu, Lei and Ai, Bo and Niu, Yong and Han, Zhu and Wang, Ning and Ma, Zhangfeng and Xiong, Lei},
	journal={IEEE Transactions on Communications}, 
	title={{STAR-RIS} Assisted Train-to-Ground Communications in Space-Air-Ground Integrated Networks}, 
	year={2025},
	month={Nov.},
	volume={73},
	number={11},
	pages={12396-12412},
	doi={10.1109/TCOMM.2025.3573456}}

@ARTICLE{10970731,
	author={Gong, Lizeng and Chen, Quan and Yang, Lei and Yin, Zhenglong and Wang, Yi},
	journal={IEEE Internet of Things Journal}, 
	title={Autonomous Traffic Prediction for {LEO} Satellite-Based {IoT} Based on Satellite Spatiotemporal Features Mapping}, 
	year={2025},
	month={Jul.},
	volume={12},
	number={14},
	pages={27021-27032},
	doi={10.1109/JIOT.2025.3562631}}

@ARTICLE{11152704,
	author={Xiao, Zhu and Wang, Rui and Bai, Jing and Li, Tong and Zhang, Shiyuan and Li, Keqin and Han, Zhu},
	title={An {LLM}-Enhanced Conditional Diffusion Model for Mobile Traffic Prediction}, 
	journal={IEEE Communications Magazine}, 
	volume={63},
	number={9},
	year={2025},
	month={Sep.},
	pages={60-67},
	doi={10.1109/MCOM.001.2400779}}

\end{document}